\documentclass[linenumber]{aastex631}

\usepackage{amsmath}
\usepackage{xcolor}
\shorttitle{Minimal Harmonics}
\shortauthors{Dipanshu et al.}
\graphicspath{{./}{figures/}}
\begin{document}

\title{Capturing Statistical Isotropy violation with generalized Isotropic Angular Correlation Functions of CMB Anisotropy}

\correspondingauthor{Dipanshu }

\author[0000-0002-9353-0910 ]{Dipanshu}
\affiliation{Department of Physics, Indian Institute of Science Education and Research, Pune 411008, India}
\affiliation{Raman Research Institute, Bangalore 560080, India}
\email{garg.dipanshu@students.iiserpune.ac.in} 
\author [0000-0003-3764-8102] {Tarun Souradeep}
\affiliation{Department of Physics, Indian Institute of Science Education and Research, Pune 411008, India}\affiliation{Raman Research Institute, Bangalore 560080, India}
\affiliation{Inter University Centre for Astronomy and Astrophysics, Post Bag 4, Ganeshkhind, Pune-411007,
India}

\author[0000-0002-4777-5044]{Shriya Hirve}
\affiliation{Department of Physics, Indian Institute of Science Education and Research, Pune 411008, India}
\affiliation{Department of Physics and Astronomy, Louisiana State University, Baton Rouge, LA 70803, USA}
\begin{abstract}
The exquisitely measured maps of fluctuations in the Cosmic Microwave Background (CMB) present the possibility to test the principle of Statistical Isotropy (SI) of the Universe through systematic observable measures for non-Statistical Isotropy (nSI) features in the data.
Recent measurements of the CMB temperature field provide  tantalizing evidence of the deviation from SI. A systematic approach based on strong mathematical formulation allows any nSI feature to be traced to known physical effects or observational artefacts. Unexplained nSI features could have immense cosmological ramifications for the standard model of cosmology. BipoSH (Bipolar Spherical Harmonics) provides a general formalism for quantifying the departure from statistical isotropy for a field on a 2D sphere. We adopt a known reduction of the BipoSH functions, dubbed Minimal Harmonics {\citep{Manakov_1996}}. We demonstrate that this reduction technique of BipoSH leads to a new generalized set of isotropic angular correlation functions (mBipoSH) that are observable quantifications of nSI features in a sky map. We show that any nSI feature in the CMB map captured by BipoSH at the bipolar multiple $L$ with projection $M$ can be studied by $(L+1)$ mBipoSH angular correlation functions in case of even parity and by $L$ functions in case of odd parity. We present in this letter a novel observable quantification of deviation from statistical isotropy in terms of generalized angular correlation functions that are compact and complementary to the BipoSH spectra that generalize angular power spectrum CMB fluctuations.

\end{abstract}


\section{Introduction} \label{sec:intro}

The CMB anisotropy measurements by WMAP \citep{WMAP:2008ydk} and Planck \citep{Planck:2018nkj} space missions have ushered in the precision era of cosmology. Precise measurements enable cosmologists to pose queries  beyond the statistically isotropic two-point correlation function predicted by the fundamental assumption of homogeneity and isotropy based on the cosmological principle. Current observations are in good agreement with CMB temperature anisotropies being Gaussian \citep{Planck:2018nkj}. In such a case, all the information encoded in the CMB temperature field can be specified by a two-point correlation function. The WMAP and Planck collaboration data release claimed significant deviation from SI in CMB maps. BipoSH provides an elegant and general formalism for the two-point correlation function for a random field on a 2-sphere, where the statistical isotropy part is just a subset in BipoSH basis \citep{Hajian_2003}. In this letter, we extend the BipoSH formalism to a new angle dependent irreducible representation to be applicable in the real (angular) space instead of harmonic basis. Departures from statistical isotropy can have its roots in known physical effects, and observational artefacts. Some known effects include Doppler boost, Weak lensing of CMB photons by large scale structure, and systematics such as non-circular beam have been studied in the BipoSH representation \citep{PhysRevD.70.103002,PhysRevD.81.083012,PhysRevD.89.083005,PhysRevD.91.043501}. With upcoming missions with great precision, the study of SI violation has far-reaching implications in cosmology. Hence it is important to study crucial signatures of departure from statistical isotropy using an appropriate mathematical construct.

\section{BIPOSH FORMALISM} \label{sec:style}
BipoSH (Bipolar Spherical Harmonics) provides a general formalism
for quantifying the departure from the statistical isotropy of CMB temperature field.
Bipolar Spherical Harmonics form a complete and orthonormal basis in $\mathbf{S}^2 \times\mathbf{S}^2$ and thus have the bidirectional dependence. The most general two-point correlation function for a field defined on the sphere can be obtained in terms of BipoSH basis as
  \begin{equation}\label{eq1}
        C\left(\hat{n}_{1}, \hat{n}_{2}\right)=\sum_{L, M, l_{1}, l_{2} } A_{l_{1} l_{2}}^{L M}
        \left\{Y_{l_{1}}\left(\hat{n}_{1}\right) \otimes Y_{l_{2}}\left(\hat{n}_{2}\right)\right\}_{L M}
        \end{equation}
where $A_{l_{1} l_{2}}^{L M}$ are BipoSH coefficients and  $\{Y_{l_{1}}\left(\hat{n}_{1}\right) \otimes Y_{l_{2}}\left(\hat{n}_{2}\right)\}_{L M} $ are Bipolar Spherical Harmonic (BipoSH) functions. BipoSH functions are tensor product of two  spherical harmonics (SH) functions that can be expanded as
\begin{widetext}

 \begin{equation}
      \{Y_{l_{1}}\left(\hat{n}_{1}) \otimes Y_{l_{2}}\left(\hat{n}_{2}\right)\right\}_{L M}=\sum_{m_{1} m_{2}} C_{l_{1} m_{1} l_{2}-m_{2}}^{L M} Y_{l_1,m_1}(\hat n_1) Y_{l_2,m_2}(\hat n_2)
        \end{equation}
            
\end{widetext}
where are $C_{l_{1} m_{1} l_{2}-m_{2}}^{L M}$ are Clebsch Gordon (CG) coefficients. The indices of CG coefficients satisfy the triangularity conditions as $|l_1 - l_2| \leq L \leq l_1 + l_2$ and  $m_1 + m_2 = M$.

    BipoSH coefficients are the natural generalization of CMB angular power spectrum.  BipoSH coefficients carry crucial signatures of SI violation describing direction-dependent statistics of CMB sky. Since the two-point correlation function is a real measurable. BipoSH is widely used for characterizing different known sources of nSI effects and systematically probing for non-statistical isotropy from the CMB maps. The $L = 0$  condition gives the isotropic part  and higher $L$ values represent the corresponding bipolar multipole of nSI effects in the CMB Sky.  

\section{Reduction Technique for Bipolar Harmonics} \label{sec:floats}
In this section, we outline a mathematical construct for the reduction of Bipolar Harmonics as studied by \cite{Manakov_1996}. In BipoSH basis, the rank $L$ has values from 0,1,2,3.. and the internal ranks $l, l^{'}$ runs over all values from  0 to infinity for a given rank $L$ constrained by the CG coefficients. In other words, the information at given Bipolar multipole $L$ could be spread over angular spectral range $l$. We show that the reduction to minimal Bipolar Spherical Harmonics (mBipoSH) limits the spectral spread to $L$ angular correlation functions with a dependence on $\hat n_1\cdot\hat n_2$. The different mBipoSH functions represent the angle-dependent field correlation functions for the nSI map.

This above reduction follows from the justification that any irreducible tensor of rank $L$ can be constructed using $L$ vectors of its arguments. Any Bipolar Harmonic with any possible internal rank $l + l^{'}$ can be constructed using combination of $L$ Minimal Harmonics defined as
 \begin{equation}\label{eq3}
   \mathcal{Y}^{k}_{LM} (\hat n_1, \hat n_2)  =   Y^{L- k,k}_{LM} (\hat n_1, \hat n_2),\ \text{where}\ k = 0, 1 ....., L.
 \end{equation}

The above relation reduces our analysis to only a few internal ranks up to $L$. The tensor with rank $l + l^{'} \geq L$ can be written from $L$ Minimal Harmonics and coefficients depending upon $l, l^{'} $and $\theta= \cos ^{-1} (\hat n_1.\hat n_2)$. Mathematical representation of the Minimal Harmonics from \cite{Manakov_1996} can be written as 
\begin{equation}\label{eq4}
      Y^{l_1l_2}_{LM} (\hat n_1, \hat n_2) = \sum_{\lambda=\lambda_p}^{L} a_{\lambda}(l_1,l_2,L,\cos\theta)  Y^{\lambda,L+\lambda_p-\lambda}_{LM} (\hat n_1, \hat n_2)
  \end{equation}
  \noindent where
 \begin{equation}
        \lambda_p =\left\{
             \begin{array}{ll}
                0 &   \text{for even $(l_1 + l_2 - L)$ }\\
                1 &   \text{for odd  $ (l_1 + l_2 - L)$ }.
             \end{array}
             \right.
  \end{equation}The parameter $\lambda_p$ describes space inversion property of Spherical Harmonics. $\lambda_p$ carries information about the parity of BipoSH functions  as discussed in \cite{PhysRevD.83.027301}. Which specifies the BipoSH function  $Y^{l_1l_2}_{LM} $ is a tensor for even parity and pseudo tensor for odd parity. The coefficients $a_{\lambda}$ follows the symmetry relation given as
  \begin{equation}\label{eq6}
         a_{\lambda}(l_1,l_2,L,\cos\theta) =a_{L-\lambda_p-\lambda}(l_2,l_1,L,\cos\theta). 
     \end{equation}The above-transformed set of Minimal Harmonics also forms a set of the complete basis for Bipolar Spherical Harmonics. The coefficients $a_{\lambda}$ are functions of angle $\theta$ between two directions, and the BipoSH functions with dependence on higher $L$ values can be constructed using these coefficients with a finite set of Minimal Harmonics basis functions. Considering the completeness property of BipoSH functions, the final expression for  $a_{\lambda}$ can be written as
\begin{widetext}
  \begin{equation}\label{eq7}
  \begin{aligned}
  a_{\lambda}^{\lambda_p}(l_1,l_2,L,\cos\theta) &= i^{\lambda_p}(-1)^{l_2} \left[ \frac{2^{L}(2l_1+1)(2l_2+1)(2L+1)!(l_1+l_2-L)!\lambda!(L-\lambda-\lambda_p)!}{q!(L-l_1+l_2)!(L-l_2+l_1)!(2\lambda+1)!!(2L-2\lambda-2\lambda_p+1)!!} \right]      \\
        & \quad            * \sum_{t=0}^{t_{max}} (-1)^{\lambda+t}  \binom{\frac{(j-l_1)}{2}}{t}\binom{L-\lambda_p-t}{\lambda}\frac{(q-\lambda_p)!!}{(q-\lambda_p- 2t)!!} P_{j-t-\lambda}^{(L-t)}(\cos\theta)
  \end{aligned}                               
     \end{equation}
\end{widetext}
    
    \noindent where $q = l_1+l_2+ L+1$, $j= L + l_2 -\lambda_p $, and $t_{max}$ = $min$ $\left[{L-\lambda_p -\lambda, \frac{j-l_1}{2}}\right]$, $\binom{m}{n}$ is the binomial coefficient and  $P^{(m)}_n(x)$ is the $m$th derivative of the Legendre polynomial $P_n(x)$. Given that the functions follow the symmetry relation given in Eq. \eqref{eq6}. Henceforth to construct the mBipoSH coefficients, Eq. {\eqref{eq7}} is only referred for  coefficients having $\lambda \geq (\frac{L}{2})$ and the remaining coefficients can be computed using the symmetry relation for the  $a_{\lambda}$. This provides us with a complete set of $a_{\lambda}$ functions for the reduction of Bipolar Harmonics with any rank $L$. It can be further analyzed from the above set of Minimal Harmonics basis that the tensor $Y^{l_1l_2}_{LM} $ has $L+1-\lambda_p$ different basis components. Using this reduction mechanism for Bipolar Harmonics, we can  construct equivalent compact and complementary  angle-dependent mBipoSH functions for given any value of multipole $L$.

\section{Constructing Minimal BipoSH functions} \label{sec:displaymath}
Employing the mechanism of reduction of Bipolar Harmonics basis discussed in previous section, we can construct the nSI features in the CMB sky maps into angle-dependent correlation functions. The most general two-point correlation function from Eq.\eqref{eq1} and Eq.\eqref{eq4} can be expanded in the form of mBipoSH angular correlation functions as
\begin{widetext}
  \begin{equation}\label{eq8}
        C\left(\hat{n}_{1}, \hat{n}_{2}\right)=\sum_{ L, M, l_{1}, l_{2}} A_{l_{1} l_{2}}^{L M}
        \sum_{\lambda=\lambda_p}^{L} a_{\lambda}(l_1,l_2,L,\cos\theta)  Y^{\lambda,L+\lambda_p-\lambda}_{LM} (\hat n_1, \hat n_2).
        \end{equation}
Where in the case of SI, the above equation gets reduced to  $C\left(\hat{n}_{1}. \hat{n}_{2}\right) = C(\theta)$. Simplifying the above equation in the case of even parity i.e. substituting $\lambda_p = 0$, the equation reduces to
 \begin{equation}\label{eq9}
      C(\hat n_1,\hat n_2) = \sum_{L, M} \sum_{\lambda=0}^{L} \left[\sum_{l_1,l_2} A^{LM}_{l_1l_2}  a_{\lambda}(l_1,l_2,L,\cos\theta) \right] Y^{\lambda,L-\lambda}_{LM} (\hat n_1, \hat n_2).
  \end{equation}
We conclude from the above expression that the correlation function for any particular multipole $L$ can be represented with  a sum over a few Bipolar functions and referring the expression inside the bracket as minimal BipoSH coefficients. Hence, mBipoSH coefficients can be written as
 
 \begin{equation}
                 \alpha^{L,M}_{\lambda }(\cos \theta) = \sum_{l_{1} l_{2}}A_{l_1 l_2}^{L M}a_{\lambda}(l_1,l_2,L,\cos\theta),
                \end{equation}
        these are the angle-dependent coefficients in our new reduced Bipolar basis. From the definition of BipoSH coefficients in \cite{Hajian_2003}, minimal BipoSH coefficients could be expressed in terms of covariance matrix computed through CMB maps as

\begin{equation}
                 \alpha^{L,M}_{\lambda }(\cos \theta) = \sum_{l_{1} l_{2}}\sum_{m_{1} m_{2}}a_{l_{1} m_{1}} a_{l_{2} m_{2}}^{*}(-1)^{m_{2}} C_{l_{1} m_{1} l_{2}-m_{2}}^{L M}a_{\lambda}(l_1,l_2,L,\cos\theta)
                \end{equation}
             \end{widetext}   
             Since $a_{lm} s$ are measurable from the CMB temperature maps. It is crucial to study of CMB sky using mBipoSH coefficients representing real space quantities. It can be emphasized from the above relations that minimal BipoSH coefficients are a set of different $\theta$ dependent correlation functions defined for specific nSI feature in a CMB map. For $L = 0 $, we recover the isotropic two-point correlation function known as CMB angular power spectrum, which is extensively studied in cosmology literature and has been source of vast information. But angle dependent isotropic correlation function for a SI violated CMB map hasn't been studied earlier. This mathematical exercise gives an easy and compact way to study higher multipole angular correlation functions in the case of nSI CMB maps. 
             
             It can be observed from the analysis that $L+1$ different angle-dependent correlation functions completely capture the anisotropy of multipole $L$ with projection $M$ in case of even parity and $L$ correlation functions in case of odd parity.
             
  The mBipoSH functions can be summed over to construct $\cos\theta$ dependent spectrum at each point on the 2D sphere.      
             
                 \begin{equation}
      \zeta_{\lambda} (\hat{n}_1,\cos\theta) = \sum_{LM}  \alpha^{LM }_{\lambda}(\cos \theta) Y_{LM}(\hat n_1) 
     \end{equation}
These defined functions  form a basis in $\mathbf{S}^2 \times\mathbf{S}^1$ and  
can minimally represent the anisotropy pattern in the CMB map. This mathematical structure opens a new avenue for constructing and analyzing CMB temperature maps.    

\section{Illustrative Example} 
Precise measurements from WMAP and Planck have signaled various sources of SI violation in the CMB data. Our mathematical representation uses harmonic space decomposed $a_{lm}$ to compute mBipoSH coefficients. These functions represent the angle-dependent real space correlation functions for SI violated map. We study one case of nSI map due to Doppler Boost with the mBipoSH angular correlations functions that have come under discussion due to results obtained by recent experiments.
\subsection{Doppler Boost}
Our motion today with respect to the cosmic rest frame causes dipole anisotropy in the CMB  temperature and polarization fields. Doppler boost of CMB  with velocity
$(\beta \equiv |\boldsymbol v|/c= 1.23 \times 10^{-3})$ induces non-zero dipolar $(L = 1)$  signatures in the CMB maps as shown by \cite{PhysRevD.89.083005}.  
Doppler boost leads to two kinds of effects, modulation, and aberration of the CMB temperature field.

This effect is studied via the $l$ , $l$ +1 correlations   of
SH space covariance matrix, which is related to the $L = 1$ BipoSH spectra.
The BipoSH coefficients for the Doppler Boost could be written as
\begin{equation}
    \tilde{A}^{1M}_{ll+1|TT}=\beta^{1M} D^{TT}_l \frac{\Pi_{ll+1}}{\Pi_{1}}C^{1 0}_{l\, 0\, l+1\, 0},
\end{equation}
where BipoSH spectra is defined as
\begin{equation}
D^{TT}_l= \, \frac{1}{\sqrt{4\pi}}\bigg[(l+ b_\nu)C^{TT}_{l} -(l+2 -b_\nu)C^{TT}_{l+1}\bigg] 
\end{equation}
where, $ \boldsymbol{\beta}$ refers to the boost velocity vector and $b_{\nu}$ is the frequency dependent effect on Doppler boost given by
\begin{equation}
    b_{\nu}= \frac{\nu}{\nu_0}\mathrm{coth}\bigg(\frac{\nu}{2\nu_0}\bigg) -1,
\end{equation}
 the local velocity $\beta_{1M}$  defined as according to \cite{2014A&A...571A..27P} 
\begin{equation}
    \beta_{1M} = \int \boldsymbol{\beta}. \hat n Y^{*}_{1M}(\hat n) d\hat n
\end{equation}

with the notation $\Pi_{l_1\, l_2 \ldots l_n}= \sqrt{(2l_1+1)(2l_2+1)\ldots (2l_n+1)}$.
The minimal BipoSH coefficients for Doppler Boost can be constructed using the method used in the above section  as
 \begin{equation}
                 \alpha^{1M}_{\lambda }(\cos \theta) = \sum_{l}A_{l l +1}^{1 M  }a_{\lambda}(l,l+1,1,\cos\theta).
                \end{equation}
The above reduction process for the Doppler Boost  $(L = 1)$  can be simplified in terms of the correlation function as
   \begin{equation}\label{eq17}
            C\left(\hat{n}_{1}, \hat{n}_{2}\right) = C_{nSI}(\hat{n}_{1}, \hat{n}_{2}) +  C_{SI}(\cos \theta).
        \end{equation}
    Where the nSI correlation function corresponding to Doppler Boost  $(L = 1)$ along the $ \boldsymbol{\beta}$ direction  in terms of angular correlation functions as 
   \begin{equation}\label{eq18}
            C_{nSI}\left(\hat{n}_{1}, \hat{n}_{2}\right) = \sum_{\lambda = 0}^{1}C_{\lambda}(\cos \theta) f_{\lambda}\left(\hat{n}_{1}, \hat{n}_{2}\right),
        \end{equation}
 
 where $C_{\lambda}(\cos \theta)$ represents the two different angular correlation functions for the Doppler-boosted CMB temperature map and $f_{\lambda}\left(\hat{n}_{1}, \hat{n}_{2}\right)$ are the functions of directions, which for the $L=1$ case, are $[\hat{n}_{1}]_{10} = \hat{n}_{1} .\hat{d}$ and $[\hat{n}_{2}]_{10}=  \hat{n}_{2} .\hat{d}$ respectively, the indices representing the zeroth component of rank 1 tensor, specifying the projection of the respective vector along the ($\hat{d}$) boost direction.
 
  Simplifying the equation becomes 
  \begin{equation}
      C_{nSI}\left(\hat{n}_{1}, \hat{n}_{2}\right) =  C_0 (\cos\theta)  \hat{n_1}. \hat{d} + C_1 (\cos\theta)  \hat{n_1}. \hat{d}
  \end{equation}
 
 The explicit form of these functions can be written as
\begin{widetext}
 \begin{equation}
    C_0(\theta)  = \sum_{l}D^{TT}_l \frac{\Pi_{ll+1}}{\Pi_{1}}C^{1 0}_{l\, 0\, l+1\, 0} \left[\frac{-\sqrt{3}}{4 \pi \sqrt{l+1}}(-1)^{l}P_{l}^{(1)}(\cos \theta)\right]
                \end{equation}
 \begin{equation}
  C_1(\theta)   = \sum_{l}D^{TT}_l \frac{\Pi_{ll+1}}{\Pi_{1}}C^{1 0}_{l\, 0\, l+1\, 0} \left[\frac{-\sqrt{3}}{4 \pi \sqrt{l+1}}(-1)^{l+1}P_{l+1}^{(1)}(\cos \theta)\right].
                \end{equation}
\end{widetext}

Naively, we can see both angular correlation functions depend upon the first derivative of Legendre polynomials $P_{\ell}^{(1)}(\cos \theta)$. This attributes to the generalization of well-studied statistical isotropic correlation function $C_{SI}(\cos \theta)$. Noting that the magnitude of both correlation functions is not exactly the same, they differ a little as their mathematical expression suggests.

Figure 1 represents the theoretical and simulated plots of angular correlation functions for the CMB map having anisotropy corresponding to $L=1$ Doppler Boost. 

\begin{figure}[h]\label{fig}
\centering
\includegraphics[width=3.2in,keepaspectratio=true]{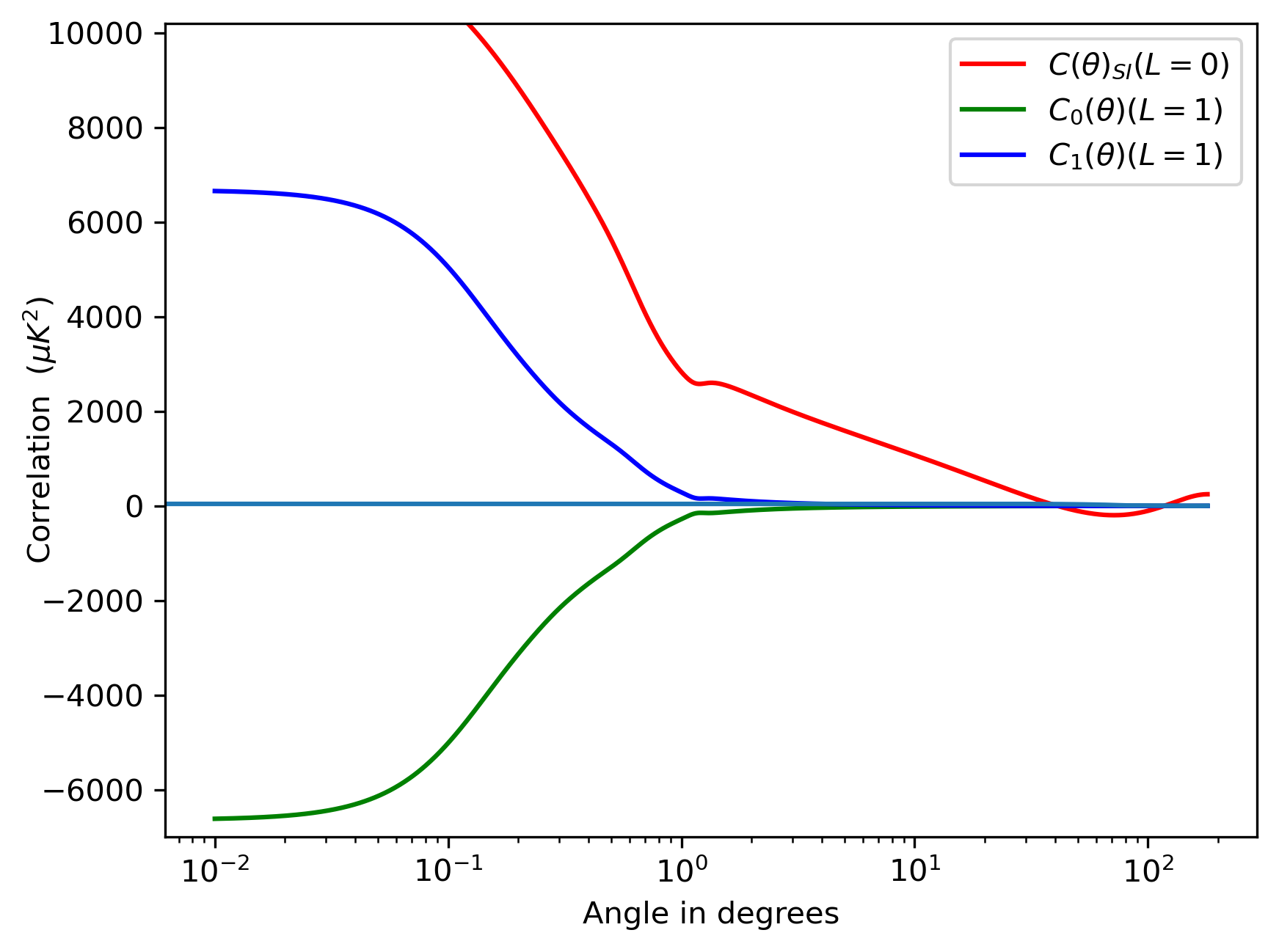}
\includegraphics[width=3.2in,keepaspectratio=true]{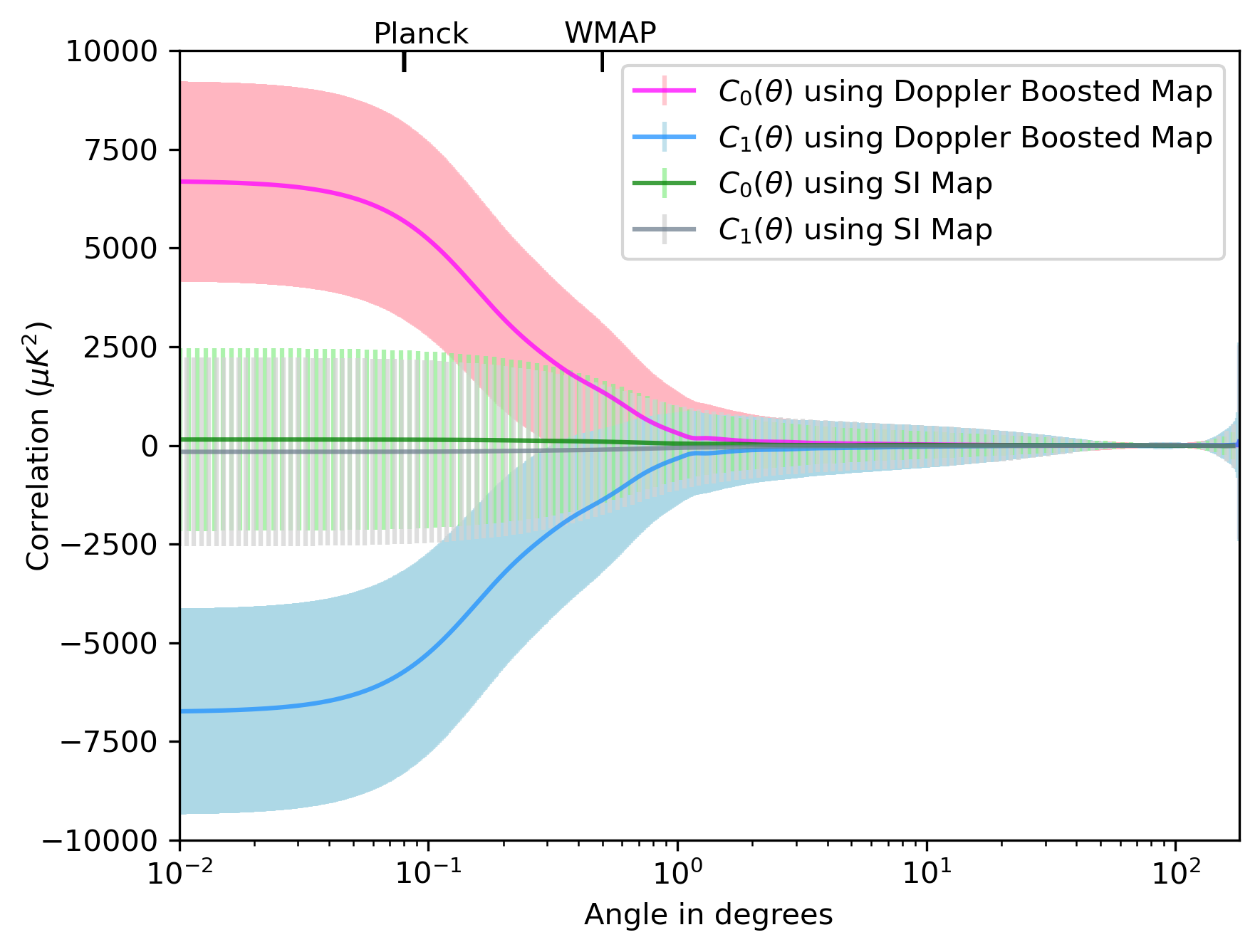}
\caption{The left panel displays theoretical correlation function $C_{SI}(\theta)$ plot for Isotropic $L=0$ part, $C_0(\theta)$ and $C_1(\theta)$ correlation functions for Doppler Boost with $b_\nu =3 $ for $\nu= 217$ GHz  with $\beta= 1.23 \times10^{-3}$ using the best-fit $\Lambda CDM$ $C_l^{TT}$ generated from CAMB \citep{2011ascl.soft02026L}. The right panel presents the  $C_0(\theta)$ and $C_1(\theta)$ plots obtained from 1000 realizations of Doppler Boosted maps using the same parameters.}\label{fig:e1} 
\end{figure}
The above plots show that temperature fluctuations in a Doppler-boosted map are correlated at very small angles only. We conclude that correlation length in fluctuations in such a case is non-zero for an angle less than $1^{\circ}$. Right panel in Figure 1 displays an error bar plot for 1000 simulated maps with such nSI effect. From the plot, we conclude that Planck mission \citep{2014A&A...571A..27P}  with much improved resolution could manifest the doppler boost velocity signal with greater efficiency as compared to WMAP \citep{hinshaw2013nine} due to signal strength relevant at very small angles only. We used the CoNIGS code discussed in \cite{Mukherjee:2013kga} for generating nSI maps for Doppler boost.

\subsection{Estimation}

We discuss here the estimation of signal strength of the source of nSI effects using the real space angular correlation mBipoSH functions.
The estimation of the mBipoSH functions could be written as
\begin{equation}\label{eqn}
\hat{\alpha}^{LM}_{\lambda}(\cos \theta)   = {\alpha}^{LM}_{\lambda}(\cos \theta) + \Gamma_{LM}G^{L}_{\lambda}(\cos \theta)\, ,
\end{equation}

where, $\hat{\alpha}^{LM}_{\lambda}(\cos \theta) $ is the observed mBipoSH coefficient  and ${\alpha}^{LM}_{\lambda}(\cos \theta) $ is the mBipoSH coefficient for a SI field, which on average over an ensemble is zero for $L \neq 0$. $\Gamma_{LM}G^{L}_{\lambda}$ is the source of SI violation with $G^{L}_{\lambda}(\cos \theta)$ as the shape factor and $\Gamma_{LM}$ is the signal strength of nSI effects like weak lensing, doppler boost, etc. This estimator is an extension of the estimator defined by \cite{hu2002mass}, \cite{hanson2009lensing} for angle-dependent real space mBipoSH coefficients.

For the the doppler boost case, to measure the $\beta_{1M}$ , we define estimator, $\hat \beta_{LM}$ for $L=1$ as

\begin{align}\label{eqnc2a}
\hat \beta_{1M}  = \sum_{\theta} \frac{{\alpha}^{1M}_{\lambda}(\cos \theta)}{{G}^{1}_{\lambda}(\cos \theta)} + \beta_{1M}\, .
\end{align}

Where, shape factor for $\lambda=1$ is defined as
\begin{equation}
  {G}^{1}_{1}(\cos\theta) =   \sum_{l}\frac{\Pi_{l, l+1}}{\sqrt{12 \pi}}\left[\left(l+b\right) C_{l}-\left(l+2-b\right) C_{l+1}\right] C_{l 0 l+10}^{10}\left[\frac{-\sqrt{3}}{4 \pi \sqrt{l+1}}(-1)^{l}P_{l}^{(1)}(\cos \theta)\right]
\end{equation}   
To arrive at the minimum
variance estimator,  using the appropriate weights, we can write

\begin{equation}\label{eqnc4}
\hat \beta_{1M}  =  \sum_{\theta}w^1_{\lambda}(\cos \theta) \frac{{\alpha}^{1M}_{\lambda}(\cos \theta)}{G^{1}_{\lambda}(\cos \theta)} + \beta_{1M}\, ,
\end{equation}
where $w^1_{\lambda}(\cos \theta)$ are the weights such that $\sum_{\theta} w^L_{\lambda}(\cos \theta) =1$. These weight factors should be chosen such that it minimizes the reconstruction noise.
This presents us with a unique estimator for the nSI effects.
\section{Discussions} 
In this letter, we proposed a natural generalization of the well-known angular correlation function that can capture non-Statistical Isotropy in sky maps of the CMB temperature fluctuations. We invoke a reduction technique for  Bipolar Spherical Harmonics that lead to new measures called Minimal Harmonics. These new measures depict the real space angular correlation functions from the nSI sky termed as mBipoSH coefficients. In the limits of SI, we recover the well-studied statistical isotropic correlation function of the CMB data.

As an illustrative example, we present the exact relations having mBipoSH angular correlations for the popular effect of nSI feature as doppler boost. We systematically construct mBipoSH functions for this case, concluding in such case of nSI effect, temperature fluctuations have correlation length at very small angles only. Through this systematic study, we have proposed that this method can be used to quantify various other signatures of SI violation in CMB data.

\begin{acknowledgments}
DG would like to thank Debabrata Adak, Rajorshi Chandra, Ritam Pal, Sayan Saha, and Shabbir Shaikh for helpful discussions during the course of this project. We also acknowledge the use of Healpix \citep{gorski2005healpix}, CoNIGS Code \citep{Mukherjee:2013kga}, BipoSH code with various contributions mentioned in
\citep{das2019si}, CAMB \citep{2011ascl.soft02026L}, Numpy \citep{harris2020array}, Scipy \citep{2020SciPy-NMeth}, Matplotlib \citep{Hunter:2007}. The work of D.G is
supported by CSIR-SRF Fellowship. S.H is partially supported by DST-Inspire fellowship.
\end{acknowledgments}

\nocite{*}
\bibliography{sample631}{}
\bibliographystyle{aasjournal}

\end{document}